# Phase formation of polycrystalline $MgB_2$ at low temperature using nanometer Mg powder

Chinping Chen [1], Zeng-jun Zhou[1], Xing-guo Li[2], Jun Xu[1], Yu-hao Wang[1], Yong-zhong Wang[1], and Qing-rong Feng[1*]

1. Department of Physics and State Key Laboratory of Artificial Microstructure and Mesoscopic Physics, Peking University, Beijing 100871, P.R. China

2. Department of Chemistry, Peking University, Beijing, 100871, P.R. China

Abstract

The $MgB_2$ superconductor synthesized in a flowing argon atmosphere using nanometer magnesium powder as the raw materials, denoted as Nano-$MgB_2$, has been studied by the technique of *in-situ* high temperature resistance measurement (HT-RT measurement). The $MgB_2$ phase is identified to form within the temperature range of 430 to 490 °C, which is much lower than that with the $MgB_2$ sample fabricated in the same gas environment using the micron size magnesium powder, denoted as Micro-$MgB_2$, reported previously. The sample density of the Nano-$MgB_2$ reaches 1.7 g/cm$^3$ with a crystal porosity structure less than a µm, as determined by the scanning electron microscope (SEM) images, while the Micro-$MgB_2$ has a much more porous structure with corresponding density of 1.0 g/cm$^3$. This indicates that the Mg raw particle size, besides the sintering temperature, is a crucial factor for the formation of high density $MgB_2$ sample, even at the temperature much lower than that of the Mg melting, ~ 650 °C. The X-ray diffraction (XRD) pattern shows a good $MgB_2$ phase with small amount of MgO and Mg and the transition temperature, $T_C$, of the Nano-$MgB_2$ was determined as 39 K by the temperature dependent magnetization measurement (M-T), indicating the existence of a good superconducting property.

Corresponding author: Qing-rong Feng

E-mail: qrfeng@pku.edu.cn

**Introduction**

Since the discovery of the superconductivity in the simple binary metallic compound of $MgB_2$ at 40 K[1], a lot of activities both in the application aspects and in the physics interests have been stimulated in the research community. In terms of basic interests, many physical properties of this material have been under intensive investigations[2-3]. Application-wise, $MgB_2$ is very promising to replace the Nb-based materials in the near future because of its high operating temperature, between 10 ~ 35 K, with a possible critical field reaching as high as 20 Tesla at 20 K[4]. The low cost in the raw materials for industrial production is also another advantages for industrial application. The art of sample preparation is therefore important. There are already intensive investigations on the fabrication techniques for making samples in the forms of film[5], tape[6], wire[6-7] and bulk with good superconducting properties. Among these, the technique for the production of bulk sample is perhaps the least complicated one. Nonetheless, the difficulties remain in avoiding the escape of the Mg vapor in order to obtain samples of good superconducting properties by a simple and effective method. Hence, it becomes an important issue to explore the various parameters for the production of bulk $MgB_2$.

In the previous reports, the effect of furnace environment conditions, both in the vacuum [8] and in the flowing argon atmosphere[9], has been investigated using micron-sized Mg powders. According to the signature of the $MgB_2$ phase formation, which is a bump followed by a sharp drop in the HT-RT curve, the phase formation occurs at 646 °C for the $MgB_2$ sample prepared in the vacuum conditions, and at a lower temperature of 537 °C for that prepared in the argon atmosphere. It remains of interests, however, to explore the effects of the raw Mg powder size. In the present work, the technique of the in-situ resistance measurements was applied to study the $MgB_2$ phase formation using nanometer-sized Mg powder in the flowing argon atmosphere. The sample was characterized by SEM observation while the superconducting transition temperature determined by the M-T measurement.

**Experiment**

Nano-$MgB_2$ samples were prepared by the solid state reaction method using boron powder

( 99.999%, grain size ≤ 1 μm) and nanometer magnesium powder (grain size ≤ 40 nm) as the raw materials. They were mixed thoroughly, and then pressed into a box made of $MgB_2$ embryo. The sample was then put into the furnace in the flowing argon conditions at ambient pressure for the heat treatment. Three stages of the heat processes were classified, according to the temperature condition of the furnace, as the rising, the holding and the cooling temperature stages. The temperature increased at a rate of 400 °C per hour, from room temperature rising to 800 °C and was then held constant for a duration of two hours and then freely cooled down to room temperature in the furnace. The 4-probe technique was applied for the in-situ HT-RT measurements during the three temperature stages. In addition to HT-RT measurement, the other characterization techniques were applied to study the samples before and after the sintering. The X-ray diffraction (XRD) spectra for the $MgB_2$ and the embryo samples before and after the sintering were taken by a Philip x'pert X-ray diffractometer. The SEM images were taken by the Stara focus ion beam (FIB) electron microscope after the furnace treatment, while the ZFC magnetization measurement was carried out at a background field of 50 Oe using a SQUID magnetometer (Quantum Design MPMS-XL)

**Results and Discussion**

The HT-RT curves of the Nano-$MgB_2$ sample are studied during 3 different temperature stages, the rising, the holding and the cooling ones, respectively. For the rising stage, the *in-situ* resistance of the Nano-$MgB_2$ is plotted versus the furnace temperature in Fig. 1. The resistance variation of the Micro-$MgB_2$ during the heat treatment from the previous work is also presented in Fig. 1 by the open circles in log-scale[9]. In the figure, both of the results show similar feature for the $MgB_2$ phase formation, a bump followed by a steep drop in resistance, despite that there exist large differences in the magnitude of the resistance and in the temperature dependent behavior before the appearance of the $MgB_2$ phase. The Nano-$MgB_2$ sample exhibits metallic nature in the HT-RT behavior with a small resistance, in the mΩ range, while the resistance of the Micro-$MgB_2$ is very large, on the order of MΩ, and drops by several orders of magnitude with increasing temperature, showing non-metallic feature. The relative particle size of the raw Mg and the B powders is an important factor in the HT-RT behavior during the initial furnace stage. The typical grain size of the boron powder is roughly 1 μm, which is small relative to 0.1∼0.2 mm of the Mg powder for the Micro-MgB$_2$ sample and large relative to 40 nm for the Nano-MgB$_2$. For the mixture of two different particles with sizes differing by more than an order of magnitude, the nature of the conducting path is usually attributed to the small-sized particles, since they easily form a complete surrounding layer around the large ones to constitute a continuous passage throughout the sample. Hence the nano-sized Mg grains constitute conducting paths to exhibit metallic conduction behavior before the formation of the MgB$_2$ phase, and the micro-MgB$_2$ shows the opposite feature. According to the previous reports[10], the MgB$_2$ superconducting phase showing up in a macroscopic scale is easily identified by the sudden drop in the HT-RT curve. This occurs from 430 to 490 °C for the Nano-MgB$_2$ and from 530 to 610 °C for the Micro-MgB$_2$, see Fig. 1. It indicates that the phase formation temperature of Nano-MgB$_2$ is lower than that of Micro-MgB$_2$ by about 100 °C.

The furnace temperature increased continuously to about 800 °C and was then kept constant for about 2 hours for the heat treatment. This is the optimum sintering temperature for the MgB$_2$[10]. The furnace temperature and the corresponding resistance of the sample are plotted against the running time in Fig. 2. There is no dramatic change in the resistance during this temperature-holding period. This indicates that the MgB$_2$ phase is very stable at the temperature applied. The small variation in resistance, which is positively correlated with the furnace temperature, reflects the temperature behavior of the MgB$_2$ superconductor. In the cooling stage, the resistance of the Nano-MgB$_2$ decreases with the furnace temperature down to the value of $4.5 \times 10^{-4}$ Ω at room temperature. This is roughly the same magnitude as that of the Micro-MgB$_2$ fabricated in the vacuum and in the argon atmosphere conditions.

The SEM images for the Nano-MgB$_2$ and the Micro-MgB$_2$ are displayed in Fig. 3(a) and (b), respectively. Both are shown in the same scale. The characteristic length for the cavity structure is barely visible, estimated on the level of 1 μm for the Nano-MgB$_2$, Fig. 3(a), while it is on the order of 100 μm or more for the Micro-MgB$_2$, Fig. 3(b). The corresponding densities are 1.7 g/cm$^3$ and 1.0 g/cm$^3$, respectively. The cavity size is evidently correlated with the grain size of the raw Mg powder. In order to fabricate samples of higher density from the elements, the small Mg powder size is therefore necessary. The XRD spectrum is shown in Fig. 4. Most of the peaks are indexed to the $MgB_2$ phase with the existence of MgO and elemental Mg. The inset of Fig. 4 shows a field-cooled (FC) and zero-field-cooled (ZFC) M-T measurement from 5 to 50 K. The transition temperature, $T_C$, is about 39 K, indicating that the sample has a good property of superconductivity.

**Conclusions**

The technique of the in-situ resistance measurement was applied to investigate the fabrication conditions of the polycrystalline $MgB_2$ superconductor using nanometer-sized raw Mg powder. The phase formation occurred at low furnace temperature, from 430 to 490 °C, in the flowing argon atmosphere. Yet, the sample density reaches 1.7 g/cm$^3$, higher than that of 1.0 g/cm$^3$ produced using micron-sized Mg powder in the same gas condition. The M-T measurement shows a good superconducting property of the sample with the transition temperature at about 39 K. The results of the present work indicate that the formation temperature of the $MgB_2$ phase and the sample density strongly depend on the raw Mg particle size, in comparison with that reported previously [8]. To date, this is the lowest phase formation temperature ever reported for the fabrication of bulk $MgB_2$ sample prepared directly from the elements.

**Acknowledgement**

This research was a part project of the Department of Physics of Peking University, and is supported by the Center for Research and Development of Superconductivity in China under contract No. BKBRSF-G1999064602.

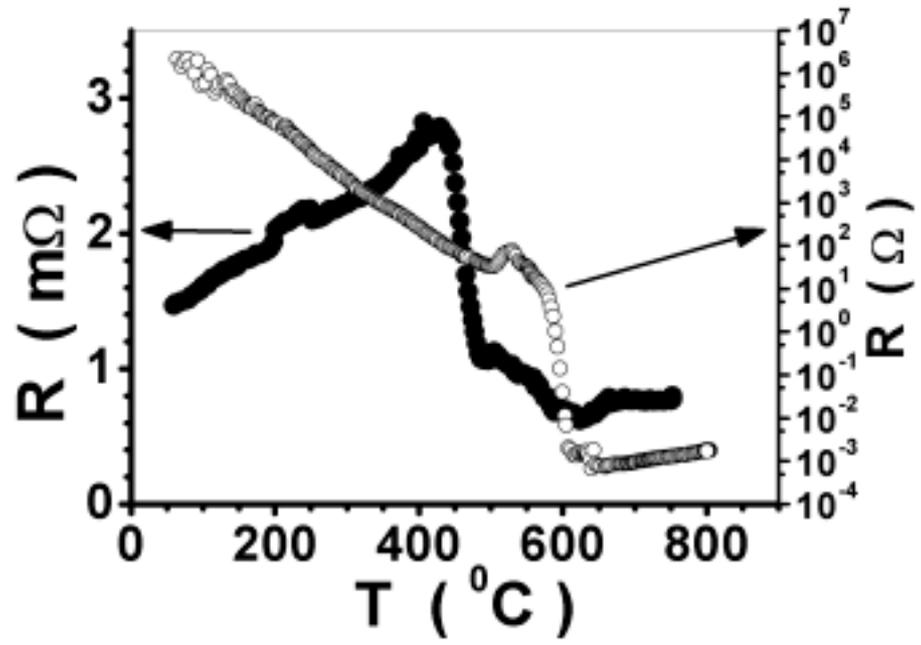

Fig. 1. HT-RT results for the Nano-MgB$_2$ in solid circles and the Micro-MgB$_2$ in open ones.

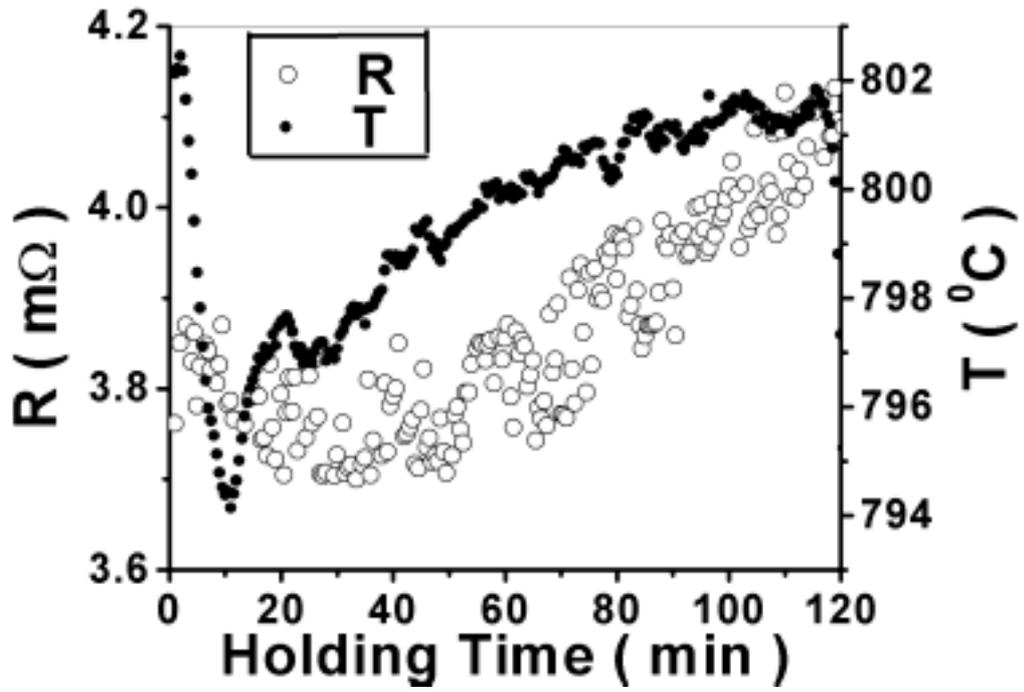

Fig. 2　Variation of the resistance, R, and the furnace temperature, T, with the furnace running time in unit of minute.

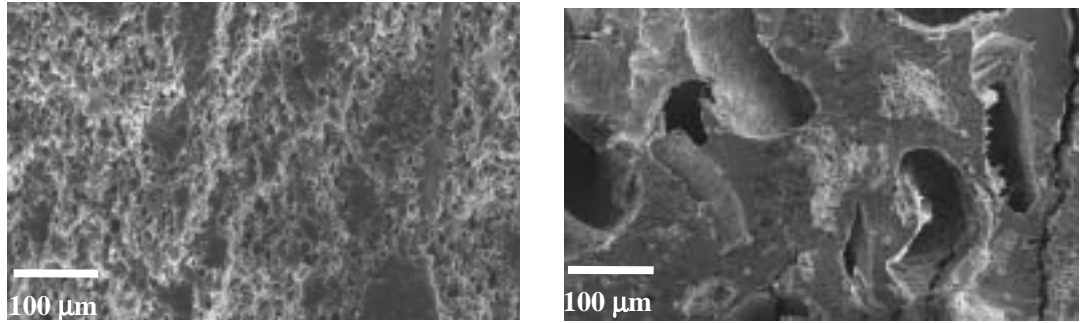

(a)                                        (b)

Fig.3. SEM images for (a) Nano-$MgB_2$ and (b) Micro-$MgB_2$ samples. Both samples were heat-treated in the flowing argon atmosphere. The Nano-$MgB_2$ shows a denser structure than the Micro-$MgB_2$ does. The densities are 1.7 and 1.0 g/cm$^3$ for (a) and (b), respectively.

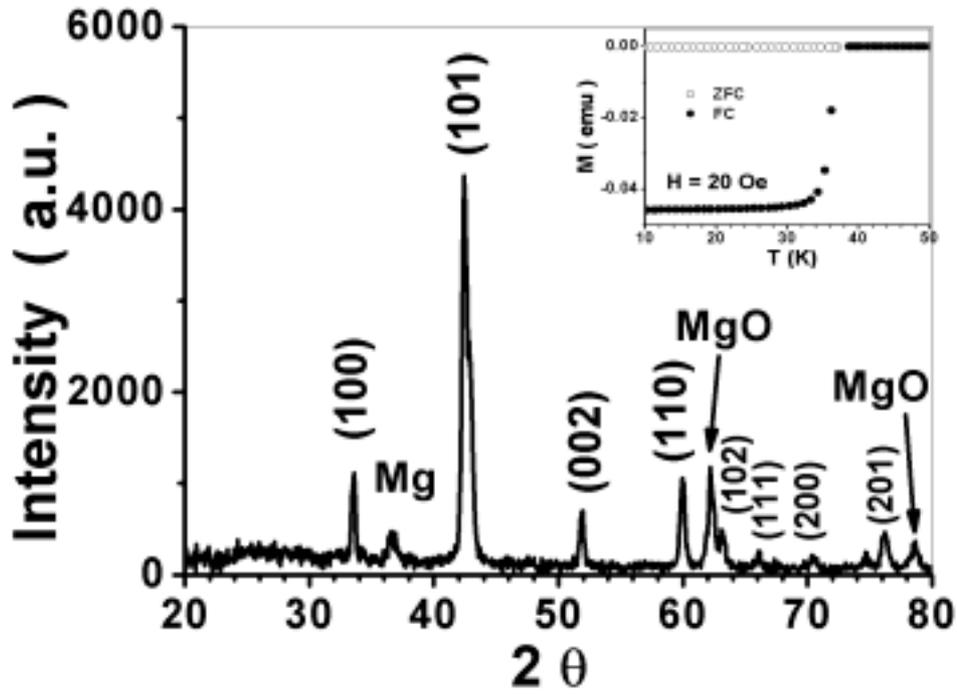

Fig. 4 XRD spectrum for the Nano-MgB$_2$. Most of the peaks can be indexed to the MgB$_2$ crystal phase with impurities of MgO and Mg. The inset is the result of FC-ZFC dc magnetization taken at applied field of 20 Oe. It shows the superconducting transition temperature at about 39 K.